\documentclass[twocolumn,prl,showpacs]{revtex4}
\usepackage{graphicx}

\begin{document}
\title{Nonlinear Band Gap Transmission in Optical Waveguide Arrays}

\author{Ramaz Khomeriki} 
\email{khomeriki@hotmail.com}
\affiliation{Department of Physics, Tbilisi State University, 3 Chavchavadze avenue, 
Tbilisi 0128, Republic of Georgia}

\date{\today}

\begin{abstract}
The effect of nonlinear transmission in coupled optical waveguide arrays is theoretically investigated via numerical simulations on the corresponding model equations. The realistic experimental setup is suggested injecting the beam in a single boundary waveguide, linear refractive index of which ($n_0$) is larger than one ($n$) of other identical waveguides in the array. Particularly, the effect holds if $\omega(n_0-n)/c>2Q$, where $Q$ is a linear coupling constant between array waveguides, $\omega$ is a carrier wave frequency and $c$ is a light velocity. Making numerical experiments in case of discrete nonlinear Schr\"odinger equation it is shown that the energy transfers from the boundary waveguide to the waveguide array above certain threshold intensity of the injected beam. This effect is explained by means of the creation and propagation of gap solitons in full analogy with the similar phenomenon of nonlinear supratransmission [F. Geniet, J. Leon, PRL, {\bf 89}, 134102, (2002)] in case of discrete sine-Gordon lattice.


\end{abstract}

\maketitle

Nonlinear phenomena in large diversity of physical systems have a close relation with each other because the nonlinear dynamics can be described only within few theoretical models \cite{dodd}. Thus there exists a possibility to predict novel effects in realistic physical systems via modeling similar processes in simple hypothetical systems, namely chains of coupled nonlinear oscillators could be used for this purpose. For instance, as recently has been discovered by Geniet and Leon \cite{leon} {\it nonlinear supratransmission} phenomenon takes place in discrete sine-Gordon lattice, this means that driving harmonically and continuously one end of the lattice with frequencies within a band gap, there is no energy flow through the lattice for low amplitude driving, while above definite driving amplitude threshold a sudden energy flow takes place. This nontrivial effect has been explained by means of the direct soliton creation at the end of the lattice, in other words the sudden energy flow occurs when the driving adjusts the internal oscillations of the localized object. It was also noted there the possibility of the existence of similar mechanism of gap soliton generation in photonic band gap materials. It should be especially mentioned that nonlinear supratransmission has been detected not only making numerical simulations for model system of discrete sine-Gordon lattice, but it is also experimentally realized on a mechanical pendulums chain driven at one end at band gap frequencies \cite{leon}. 

\begin{figure}[t]
\begin{center}\leavevmode \hspace{-0.5cm}
\includegraphics[width=1.1\linewidth,clip]{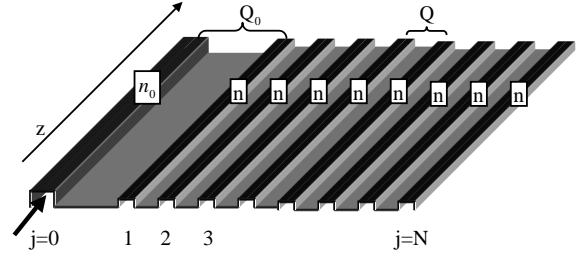}
\end{center}
\vspace{-2cm}
\caption{Suggested experimental setup: The beam is injected into the boundary waveguide numbered as $j=0$, $Q_0$ is a linear coupling between boundary and first waveguides and $Q$ is a coupling constant between the waveguides in the array. $n_0$ and $n$ are linear refractive indexes of boundary and array waveguides, respectively; $z$ is a longitudinal space dimension playing a role of time in boundary driven DNLS equation (\ref{1}).}
\label{array}
\end{figure}

The present letter aims to analyze whether the similar scenario takes place in case of discrete nonlinear Schr\"odinger (DNLS) equation and then make the predictions concerning the corresponding nonlinear processes in coupled optical waveguide arrays \cite{christo}. The experimental conditions are suggested for which optical waveguide array becomes transparent with respect to the beam injected into the single boundary waveguide if beam's intensity exceeds certain threshold (see Fig \ref{array}). This effect is opposite to the ordinary case when for low intensities (linear regime) the light injected into the single waveguide spreads to other waveguides, while in nonlinear case the light is trapped into several neighboring waveguides leading thus to the spatial discrete optical breather creation (see e.g. most recent experimental papers on the subject \cite{erice1}). It should be just mentioned that longitudinal space dimension in optical waveguide array plays a role of a time variable and one should especially care about this suggesting real experiments on waveguide arrays. 

Let us start form the consideration of boundary driven DNLS equation which could be written in the following form (j=1, $\dots$ , N):
\begin{equation}
i\frac{\partial \psi_j}{\partial z}+\psi_{j+1}+\psi_{j-1}+2|\psi_j|^2\psi_j=0; \qquad \psi_0=Ae^{i\Delta z}. \label{1}
\end{equation}
Here $z$ variable stands for the time, $\Delta$ and $A$ are driving frequency and amplitude of the boundary. Initial condition reads as $\psi_j(0)=0$ and it is supposed that the driving is turned on adiabatically, e.g. $A=A_0\bigl[1-\exp(-z/\tau)\bigr]$ in order to avoid the appearance of the perturbations from the initial shock. In simulations it is taken $\tau=10$ and nonlinear dynamics is monitored up to the time scales $10^4$, thus in stationary regime ($z\gg \tau$) one has $A=A_0$. The damping also has been applied at the right end of the waveguide array in order to suppress edge reflection. Note that in the absence of driving the sum of intensities $\sum_j^N|\psi_j|^2$ is a conserved quantity, thus the nonlinear dynamics across the array could be described via intensity flux $J_j$ through the site $j$:
\begin{equation}
J_j=i\bigl(\psi_j\psi_{j+1}^*-\psi_j^*\psi_{j+1}\bigr) \label{2}
\end{equation}
where the intensity and intensity flux at site $j$ are connected with each other via discrete continuity condition $d|\psi_j|^2/dz+\bigl(J_j-J_{j-1}\bigr)=0$.

\begin{figure}[t]
\vspace{-6cm}
\begin{center}\leavevmode 
\includegraphics[width=0.9\linewidth,clip]{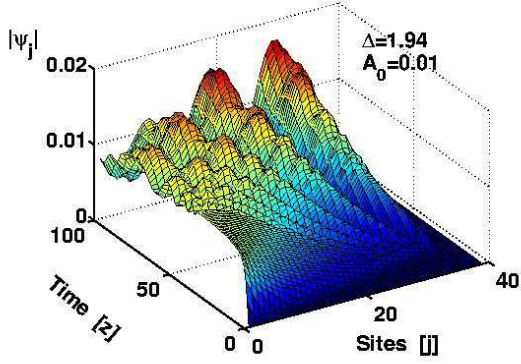}
\vspace{-0.5cm}
\end{center}
\caption{(Color online) Three dimensional plot of time evolution of boundary driven DNLS equation for inband driving frequency $\Delta=1.94$ and very small driving amplitude $A_0=0.01$. As seen the intensity transmits to remote sites.}
\label{inband}
\end{figure}

\begin{figure}[t]
\begin{center}\leavevmode 
\vspace{-4cm}
\includegraphics[width=1.1\linewidth,clip]{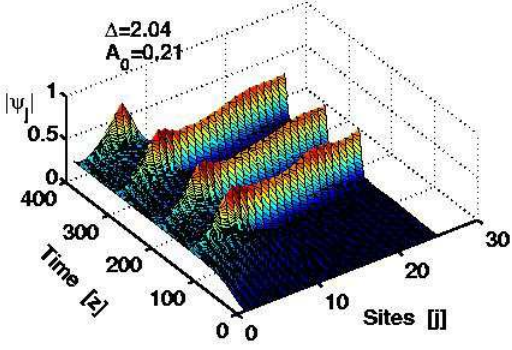}
\vspace{-5.5cm}
\end{center}
\caption{(Color online) The transmission process for the driving amplitudes above the threshold, which could be described by a train of gap solitons (\ref{3}).}
\label{bandgap2}
\end{figure}

The numerical simulations have been performed choosing different values of boundary driving parameters $\Delta$ and $A$. From the numerical experiments it follows that boundary driving leads to the perturbation of all sites if driving frequency $\Delta$ is located within the linear phonon band $-2<\Delta<2$, i.e. there is a nonzero intensity flux for any driving amplitudes (see Fig. \ref{inband}). On the other hand if the driving frequency is in upper band gap, particularly $\Delta>2$, for low driving amplitudes only several neighboring sites are excited and intensity flux to remote sites is zero. The energy starts to flow only if the driving amplitude exceeds certain threshold (see Figs. \ref{bandgap1} and \ref{bandgap2}). Note that if the driving frequencies are within a lower ban gap there is no intensity flow for any driving amplitudes. Now it is time to analyze the mechanisms for that effect.

As far as a boundary driving is applied it is natural to expect that localized solutions will excite. This statement is in full accordance with the consideration of similar process in discrete sine-Gordon type models where the {\it nonlinear supratransmission} has been discovered \cite{leon}. Thus one can assume that nonzero intensity flux will appear when boundary driving excites moving localized solutions. It is easy to derive a semi-discrete approximate envelope soliton solution substituting ansatz $\psi_j=\Psi(j)\exp\{i(\beta z-\chi j)\}$ into the DNLS equation (\ref{1}). Then assuming that the envelope $\Psi(j)$ varies smoothly along the lattice and expanding $\Psi(j\pm 1)$ about the site $j$ we get the following approximate one soliton solution (see for the details of similar derivation e.g. in Ref. \cite{rani}):
\begin{figure}[t]
\vspace{-6.5cm}
\begin{center}\leavevmode 
\includegraphics[width=1.2\linewidth,clip]{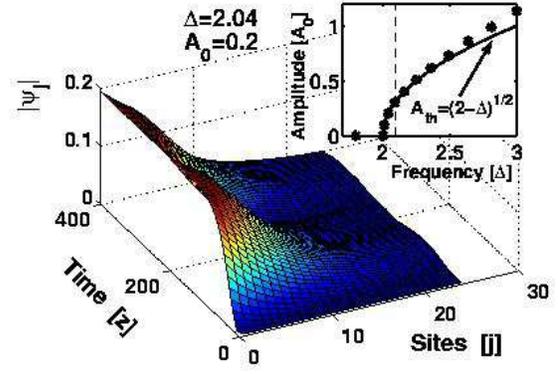}
\vspace{-4cm}
\end{center}
\caption{(Color online) The same as in Fig. \ref{inband} for band gap driving $\Delta=2.04$. For driving amplitudes below the threshold ($A_{th}=0.202$) the pattern in this graph could be described by standing breather solution (\ref{6}). Inset shows dependence of driving threshold upon the driving frequency, asterisks are results of numerical simulations on boundary driven DNLS equation (\ref{1}) and solid line represents analytical curve (\ref{5}). Dashed line divides the range of $A_0$ (left hand side) for which analytical approximate semi-discrete approach given by formulas (\ref{3}) and (\ref{6}). Moreover within the same range above threshold low amplitude semi-discrete envelope solitons participate in band gap transmission (see Fig. \ref{bandgap2}). While at the right side of the dashed line in the inset high amplitude moving breathers are excited which are further trapped by the lattice (see Fig. \ref{pin}), which leads to the suppression of band gap transmission.}
\label{bandgap1}
\end{figure}

\begin{equation}
\psi_j=\frac{|\psi_j|_{max}}{\cosh\bigl[|\psi_j|_{max}(j-Vz)\bigr]}e^{i(\beta z-\chi j)} \label{3}
\end{equation}
with a following nonlinear dispersion relation for the carrier wave of the envelope soliton ($\chi$ varies from 0 to $2\pi$) 
\begin{equation}
\beta=2\cos\chi+|\psi_j|_{max}^2, \label{4}
\end{equation}
and $V=\partial \beta/\partial \chi=-2\sin\chi$ is a soliton's group velocity. 
Note that the assumption that soliton envelope varies smoothly along the lattice puts the following restriction on the soliton amplitude $|\psi_j|_{max}^2\ll 1$. 

It is expected that intensity flux appears in the system if driving adjusts the nonlinear dispersion relation (\ref{4}), i.e. $\beta=\Delta$ and $|\psi_j|_{max}=A_0$. In other words, flux is nonzero only if one can find such $\chi$ that the following condition is fulfilled $\Delta=2\cos\chi+A_0^2$. Therefore for inband driving $-2<\Delta<2$ the nonzero flux appears even for very low driving amplitudes $A_0$, while in upper band gap 
\begin{equation}
\Delta>2 \label{delta}
\end{equation}
there exists certain amplitude threshold
\begin{equation}
A_{th}=\sqrt{\Delta-2} \label{5}
\end{equation}
below which ($A_0<A_{th}$) there is no intensity flux into the system. Instead, only several sites are excited and that pattern could be described \cite{lepri} by a static breather solution of Eq. (\ref{1}). This solution could be derived from the general one (\ref{3}) for zero velocity $V=-2\sin\chi=0$, i.e. $\chi=0$ and requiring the fulfillment of boundary condition $\psi_0=A_0\exp(i\Delta z)$ in stationary state ($z\gg \tau$):
\begin{equation}
\psi_j=\frac{\sqrt{\Delta-2}}{\cosh\bigl[(j+x_0)\sqrt{\Delta-2}\bigr]}e^{i\Delta z}, \label{6}
\end{equation}
where $x_0=\mbox{acosh}\bigl[\sqrt{\Delta-2}/A_0\bigr]/\sqrt{\Delta-2}$.

\begin{figure}[t]
\vspace{-3cm}
\begin{center}\leavevmode 
\includegraphics[width=0.8\linewidth,clip]{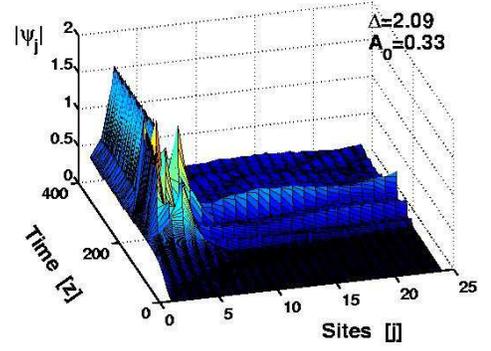}
\vspace{-2.5cm}
\end{center}
\caption{(Color online) Trapping of large amplitude moving gap soliton. For relatively large driving amplitudes the soliton starts to move but further the lattice traps it. This itself stops the transmission process.}
\label{pin}
\end{figure}

The results of numerical simulations are fully explained by the above consideration. In inset of Fig. \ref{bandgap1} it is presented the comparison between numerical experiments on Eq. (\ref{1}) and analytical formula (\ref{5}) for the driving threshold above which ($A_0>A_{th}$) a nonzero intensity flux appears in the system. For driving frequencies $\Delta$ close to 2 there is a perfect agreement, but this agreement becomes worse for larger driving frequencies. The point is that for driving frequencies sufficiently larger than 2 threshold amplitudes become comparable with unity according to the relation (\ref{5}). But for such amplitudes the continuum envelope approximation (\ref{3}) is invalid, moreover, as numerical simulations show large amplitude excitations are trapped by the lattice, (see Fig. \ref{pin}), 
as a result localizations do not move and intensity flux becomes zero. Thus the band gap transmission effect exists if there are moving solutions in the system. As a result in case of DNLS equation the discovered phenomenon is observable for driving frequencies $2<\Delta<2.09$. 

Now let us discuss how the obtained results could be applied to describe nonlinear transmission processes in the system of coupled optical waveguides. For the purpose to realize a band gap driving it is suggested (see Fig. \ref{array}) to inject a beam into boundary waveguide with linear refractive index $n_0$ larger than refractive index $n$ of other waveguides forming the array. Let us introduce a linear coupling constant between array waveguides as $Q$, while the coupling between boundary ($j=0$) and first waveguides is defined as $Q_0$. Besides that, let us suppose for simplicity that the nonlinear refractive index (kerr nonlinearity) in $j=0$ waveguide is equal to zero and onsite nonlinear refractive index in array waveguides is $D$. Thus the wave envelopes in waveguides could be described by a set of the following equations ($j=2,\dots, N$):
\begin{eqnarray}
i\frac{\partial \psi_0}{\partial z}+\frac{\omega}{c}n_0\psi_0+Q_0\psi_{1}=0 \nonumber \\
i\frac{\partial \psi_1}{\partial z}+\frac{\omega}{c}n\psi_1+Q_0\psi_{0}+Q\psi_{2}+D|\psi_1|^2\psi_1=0 \label{12}\\
i\frac{\partial \psi_j}{\partial z}+\frac{\omega}{c}n\psi_j+Q(\psi_{j+1}+\psi_{j-1})+D|\psi_j|^2\psi_j=0. \nonumber
\end{eqnarray}

where $\omega$ is a carrier wave frequency and $c$ is a light velocity. Last equation from the set (\ref{12}) is a well known representation of infinite waveguide array by the DNLS equation \cite{christo} while first two equations describe influence of boundary waveguide (with different linear refraction index) on the semi infinite array (see also Ref. \cite{kivshar}). After the appropriate rescaling 
\begin{eqnarray}
\psi_j=\psi'_je^{iz\omega n/c}\sqrt{2Q/D} \quad \mbox{for} \quad j=1,\dots, N; \qquad \label{13} \\
\psi_0=\psi'_0e^{iz\omega n/c}\bigl(Q/Q_0\bigr)\sqrt{2Q/D}; \qquad z=z'/Q \qquad \nonumber
\end{eqnarray}
Eqs. (\ref{12}) obtain simpler form ($\Delta \equiv \omega\bigl(n_0-n\bigr)/Qc$):
\begin{eqnarray}
i\frac{\partial \psi'_0}{\partial z'}+\Delta\psi'_0+\frac{Q_0^2}{Q^2}\psi'_{1}=0 \label{14} \\
i\frac{\partial \psi'_j}{\partial z'}+(\psi'_{j+1}+\psi'_{j-1})+|\psi'_j|^2\psi'_j=0. \nonumber
\end{eqnarray}
($j=1,\dots, N$)  which reduces to the boundary driven DNLS (\ref{1}) with boundary condition $\psi'_0=\psi'_0(0)\exp(i\Delta z')$ in the limit $(Q_0/Q)\rightarrow 0$ and therefore for $Q_0/Q\ll 1$ one can use the results derived for the case of boundary driven DNLS (\ref{delta}) and (\ref{5}). Particularly, for 
\begin{equation}
\Delta\equiv \omega\bigl(n_0-n\bigr)/Qc >2 \label{15}
\end{equation} 
the localized excitations (\ref{3}) or (\ref{6}) form with propagation constant located in the upper band gap $\beta=\Delta>2$. Thus if the injected intensity in the boundary is below the threshold
\begin{equation}
|\psi_0(0)|^2_{th}=\frac{2Q^3}{DQ_0^2}|\psi'_0(0)|^2_{th}\simeq \frac{2Q^2}{DQ^2_0}\left[\frac{(n_0-n)\omega}{c}-2Q\right] \label{16}
\end{equation}

one has a static breather solution (\ref{6}) and intensity flux to the array waveguides is equal to zero, while above the threshold energy transmission begins via gap solitons (\ref{3}). 
\begin{figure}[t]
\vspace{-3cm}
\begin{center}\leavevmode 
\includegraphics[width=0.9\linewidth,clip]{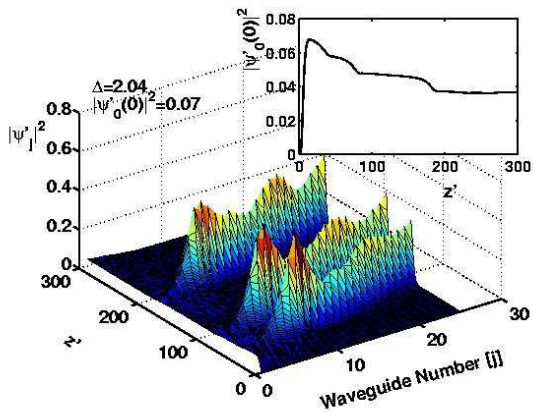}
\vspace{-3cm}
\end{center}
\caption{(Color online) Rescaled waveguide intensity $|\psi'_j|^2$ in the band gap transmission regime. Inset shows longitudinal dimension $z'$ dependence of rescaled beam intensity in boundary waveguide. Creating the solitons the intensity from the boundary waveguide is transferred to other waveguides and as a result the transmission stops when beam intensity in boundary waveguide goes below the threshold. The simulations are done for $Q_0/Q=0.1$. }
\label{guide}
\end{figure}

The above expression for the threshold (\ref{16}) becomes exact in the limit $(Q_0/Q)\rightarrow 0$. For small but nonzero $(Q_0/Q)$ the beam intensity in the boundary waveguide could be considered as almost constant quantity irrespective to the spread of energy in nonlinear band gap transmission regime, because according to the rescaling beam intensity in the boundary waveguide is $Q/Q_0$ times larger than the amplitude of propagating soliton through the array. In case when $Q_0$ becomes comparable with $Q$ almost all intensity in boundary waveguide is needed to form a soliton. As a result the intensity in boundary waveguide sharply decreases and therefore much larger threshold intensity is needed [than given by relation (\ref{16})] to develop the nonlinear transmission in band gap regime. In Fig. \ref{guide} it is presented the picture for the case $Q_0/Q=0.1$. As seen the beam intensity in boundary waveguide is above the threshold at the origin $z'=0$ and it produces gap solitons causing the band gap transmission. But this process itself causes the decrease of beam intensity in the boundary waveguide, the intensity after the creation of several solitons goes below the threshold and transmission process is not observable for large $z'$.

As it has been mentioned above nonlinear band gap transmission regime holds if one has low amplitude solitons. Large amplitude solitons tend to pin and energy transfer becomes much less effective. However, this is true only in case of DNLS equation where only onsite nonlinearities are taken into account. As shown recently \cite{erice2} considering also the terms describing also intersite nonlinearities one has moving breather solutions even at large excitations amplitudes. High intensity moving breathers have been also detected on the recent experiments \cite{erice1}. Therefore the numerical simulations have been undertaken adding to DNLS equation also terms with intersite nonlinearities. In this case the energy transfer via the moving breathers take place even for large excitation amplitudes and as a result optical transparency regime is observable in whole range of $\Delta>2$.

Summarizing it should be noted again that the novel scenario of nonlinear band gap transmission in optical waveguide arrays is predicted and simple experimental setup is suggested for its realization. Suggested experimental setup would serve also for generation of optical gap solitons propagating across the waveguide array.

Finally I would like to thank all participants of NATO advanced research workshop at Erice (Italy, 2003) for inspiring discussions. Special thank go to Yuri Kivshar, Stefano Lepri, Stefano Ruffo and Lasha Tkeshelashvili for suggestions. The work is supported by USA CRDF award No GP2-2311-TB-02.

\end{document}